\documentclass[epjST]{svjour}
\usepackage{color}
\usepackage{graphicx}
\usepackage{multirow}
\usepackage{amsmath}
\usepackage{amssymb}
\usepackage{bm}
\begin{document}

\title{Boundary interface conditions and solute trapping near the transition to diffusionless solidification}

\author{G.\,L. Buchbinder\inst{1}\fnmsep\thanks{\email{glbuchb@yahoo.com.}},  P.\,K.~Galenko\inst{2,3}}
\institute{Omsk State University, Physics Department, 644077 Omsk,
Russia  \and Ural Federal University, Theoretical and Mathematical
Physics Department, Laboratory of Multi-Scale Mathematical Modeling,
620000 Ekaterinburg, Russia \and
Friedrich-Schiller-Universit{\"{a}}t-Jena, Faculty of Physics and
Astronomy, Otto Schott Institute of Materials Research, 07743 Jena,
Germany}

\mail{correspondence author}

\abstract{The process of rapid solidification of a binary mixture is
considered in the framework of local nonequilibrium model (LNM) based on
the assumption that there is no local equilibrium  in solute
diffusion in the bulk liquid and at the solid-liquid interface. According to LNM the transition
to complete solute  trapping and diffusionless solidification occurs
at a finite interface velocity $V=V_D$, where $V_D$ is the diffusion speed
in bulk liquid. In the present work, the boundary
conditions at the phase  interface moving with the velocity $V$
close to $V_D$ ($V \lesssim V_D$) have been derived to find the
non-equilibrium  solute partition coefficient. In the high-speed region, its
comparison with the partition coefficient from the work [Phys. Rev.
E 76 (2007) 031606] is given. }

\titlerunning{--Rapid solidification...}

\maketitle


\section{Introduction}

In the present  time  the process of rapid solidification  is a well
established method  for production of the metastable materials and,
in particular, supersaturated solid solutions, which can form due to
the decrease of solute segregation at the rapidly moving
solid-liquid  interface \cite{KF92,H94,HGH07,GA18,GJ19}.
Quantitatively this effect can be characterized by the partition
coefficient $k$ defined as the ratio of solid and liquid
concentrations of the
 solute at the phase interface. The phenomenon of "solute
trapping"\, by the growing phase implies the deviation of chemical partition coefficient from its equilibrium value $k_e$ with its
increasing towards unity at large growth rates. This phenomenon has been attracting
considerable attention over of several decades both from
experimental and theoretical points of view
\cite{KF92,H94,HGH07,GA18,GJ19,W80,A82,AK88,AW86,AB94,SA94,RS94,KA95,KS00,JK04,BJ04,EC92,WB93,RB04,AW98,AA99,S95,S96,S97,GS97,G02,GD04,GH06,G07}.

In experiments with rapid solidification, very high velocities of
the phase interface can be reached such that the deviations from
local equilibrium in bulk phases and at an interface become
considerable \cite{A82,AK88,KA95,KS00,S95,GS97,BC69}. For
theoretical description of solute trapping and related phenomena
observed during rapid solidification  a number models have been
proposed \cite{W80,A82,AK88,AB94,JK04,BJ04} in which, in particular,
the deviation from local (chemical) equilibrium at solid-liquid
interface is described by the partition coefficient $k(V)$ depending
on growth velocity $V$. These models predict that the complete
solute trapping with $k = 1$ is possible only asymptotically at $V
\rightarrow \infty$.

Meantime there are  a number of the experimental works
\cite{KA95,KS00,EC92,DW60,BC65,M82} in which it has been shown that
the transition to complete solute trapping giving rase to
diffusionless solidification  occurs at substantially finite  values
of $V$. This circumstance is automatically taken into account within
the scope of the local nonequilibrium model (LNM) developed in the
works \cite{GJ19,S95,S96,S97,GS97,G02,G07}. At high growth
velocities the deviation from local equilibrium can be essential not
only at the interface  but in the bulk of the liquid phase as well.
The partition coefficient taking into account these points has been
introduced in \cite{S95,S96} and for the concentrated melt in
\cite{G07} and in the latter case it has the form
\begin{equation}\label{eq1}
k(V) = \left\{\begin{array}{ll}
\displaystyle{\frac{(1 - V^2/V^2_{D})[k_e + (1 - k_e)c_0] + V/V_{DI}}{1 - V^2/V^2_{D} + V/V_{DI}}}\,,&   V < V_{D}\\
1\,,& V \geq V_{D}\,,
\end{array}\right.
\end{equation}
where $c_0$ is the initial concentration of the melt and $V_{DI}$ is
the atom diffusive speed at the interface. The expression
(\ref{eq1}) takes naturally into account the fact that when the
growth velocity $V$  exceeds the speed $V_D$ of the
concentration disturbances  propagation in the liquid, the solute
transfer in the liquid bhas no time to occur and the transition to the
diffusionless solidification begins at the finite velocity  $V =
V_D$.

Despite the fact that the partition coefficient (\ref{eq1}) is
consistent with experimental data and MD modeling of rapid
solidification of a number of binary systems
\cite{EC92,WH90,GR07,HG08,GR09,AG17,YH11}, some theoretical
questions still remain open. In particular, according to LNM the transition to complete solute trapping at $V = V_D$ is purely a diffusion effect independent on the details of interface kinetics (i.e., independently of the collision-limited or  diffusion-limited growth mechanism exists at the atomically rough or atomically smooth faces of the crystal).
This means that in the high-speed region, $V \lesssim V_D$, the nonequilibrium
partition coefficient $k(V)$ is also likely not to experience such a
dependence and to a greater extent should be determined from the
macroscopic boundary conditions at the interface.

Bearing above, it is interesting to consider the process of rapid
solidification near the transition to a complete solute trapping,
$V\lesssim V_D$, when the state of the system out of local
equilibrium and local nonequilibrium diffusion effects play a
significant role. The  purpose of this work is to derive the
boundary conditions at the interface in this high-speed region to
analyze the nonequilibrium partition coefficient.

\section{Boundary conditions}

Let us consider a solidifying binary mixture of two species $A$ (solvent) and
$B$ (solute).  The process of nonequilibrium solidification is
accompanied by an increase in entropy. In the case of isothermal
solidification and the absence of convection in the bulk phases the
interface entropy production, $\sigma$, at the sharp interface has
the form  \cite{CC84,CC86}
\begin{equation}\label{eq2}
   T\sigma = j_A\Delta\mu_A + j_B\Delta\mu_B\,,
\end{equation}
where $T$ is the interface temperature, $\Delta\mu_i = \mu^S_i -
\mu^L_i$ and $\mu^{LS}_i$ are the chemical potentials per unit mass
of species
 $i$ $(i = A, B)$ at the liquid ($L$)
 and solid ($S$) sides of the interface. Normal   to the interface the component of the mass current of $i$-th species is defined by
\begin{equation}\label{eq3}
    j_i = \rho^{LS}_i(\bm{v}_i^{LS} - \bm {V})\cdot \bm{n},
\end{equation}
where $\bm{n}$  is the unit vector normal to the interface pointing
 into the liquid , $\bm {V}$ is the interface velocity,
$\bm{v}_i^{LS}$ is the velocity of species $i$ and $\rho^{LS}_i$ its
mass density  in the $L$ ($S$) phase.  Due to mass conservation,
each of the currets $j_i$'s is conserved across the interface.

Equality (\ref{eq2}) defines the differences of chemical potentials
$\Delta\mu_i$ as thermodynamic driving forces which cause mass
currents of $j_i$-component\footnote{The connection of  equality~(\ref{eq2}) with the local
nonequilibrium approach will be discussed below, after the formula
(\ref{eq14}).}. From a physical point of
view, it is more convenient to use  another system of independent
currents, namely  the total mass current
\begin{equation}\label{eq4}
J = j_A + j_B \,,
\end{equation}
 and the diffusion solute current of
in  each phase
\begin{equation}\label{eq5}
 J_D^{LS} = \rho^{LS}_B\bm{v}_B^{LS}\cdot \bm{n} =  (1 - C_{LS})j_B
 - C_{LS}j_A ,
\end{equation}
where $C_{LS} = \rho^{LS}_B/\rho$ is the mass concentration species
$B$  and  the density of the medium at both sides of the interface
is assumed to be equal to $\rho$.

Using (\ref{eq4}) and (\ref{eq5}) one can rewrite the production of
entropy in the form (see also \cite{CC86})
\begin{subequations}
\label{eq6}
\begin{equation}
 T\sigma = [(1 - C_L)\Delta\mu_A + C_L\Delta\mu_B]J  + (\Delta\mu_B - \Delta\mu_A)J_D^L,\label{subeq:1}
\end{equation}
\begin{equation}
\phantom{T\sigma} = [(1 - C_S)\Delta\mu_A + C_S\Delta\mu_B]J  +
(\Delta\mu_B -
 \Delta\mu_A)J_D^S.\label{subeq:2}
\end{equation}
\end{subequations}
The combination of the relations (\ref{eq6}), together with the
equality $J = j_A + j_B = -\rho V$, where $V = \bm {V}\cdot \bm{n}$,
gives the well-known boundary condition for the diffusion currents
\begin{equation}\label{eq7}
   J_D^L - J_D^S = (C_S - C_L)J = (C_L - C_S)\rho V.
\end{equation}
Taking into account that the
diffusion current can be neglected in the solid phase, $J_D^S = 0$, further will be convenient for the
production of entropy to proceed from the expression
(\ref{subeq:2}):
\begin{equation}\label{eq8}
    T\sigma = - [(1 - C_S)\Delta\mu_A + C_S\Delta\mu_B]\rho V.
\end{equation}
The Gibbs free energy change of the system for the formation of a
unit mass of solid of composition $C_S$,  $\Delta G_m = (1 -
C_S)\Delta\mu_A + C_S\Delta\mu_B$, included in equality (\ref{eq8}),
is the thermodynamic driving force causing the mass current $J = -
\rho V$. For small $V$ the linear Onsager relation follows  from
(\ref{eq8})
\begin{equation}\label{eq9}
\Delta G_m = - LV,
\end{equation}
with the positive kinetic coefficient $L > 0$ providing the positive definiteness of $\sigma$ and
$\Delta G_m < 0$. At finite $V$, the linear approximation becomes
unsuitable and the right hand side in (\ref{eq9}) should be replaced by
some nonlinear function of $V$, i.e.
\begin{eqnarray}
  \Delta G_m&=& (1 - C_S)\Delta\mu_A + C_S\Delta\mu_B = - f(V) \label{eq10}\\
 f(V)& \geqslant& 0.\nonumber
\end{eqnarray}
The form of function $f(V)$ can be set from the following
considerations. It is known \cite{T62} that the velocity of the
interface is related to the free energy change for solidification of
one mole of substance $\Delta G = M\Delta G_m$, where $M$ is the
molar mass, by the kinetic equation
\begin{equation}\label{eq11}
    V = V_0(1 - e^{ M\Delta G_m/RT}),
\end{equation}
where $V_0$ is the upper limit interface speed at $\Delta G
\rightarrow \infty$ and $R$ is  the gas constant. Comparing this
expression with (\ref{eq10}), one obtains
\begin{equation}\label{eq12}
    f(V) = - \frac{RT}{M}\ln(1 - V/V_0).
\end{equation}
Thus, the boundary conditions at the interface,
moving at an arbitrary velocity, can be represented as
\begin{eqnarray}
  (1 - C_S)\Delta\mu_A + C_S\Delta\mu_B &=& \frac{RT}{M}\ln(1 - V/V_0)\label{eq13}\\
J_D^L - J_D^S = (C_L - C_S)\rho V.&&\label{eq14}
\end{eqnarray}
It should be noted that in the derivation of the thermodynamic
equality (\ref{eq2}) no assumptions about the nature of dissipative
processes in the bulk of phases were made, so that (\ref{eq2}) and
(\ref{eq13}) are valid also in the local nonequilibrium state, for
which $J_D^L$ satisfies the Maxwell-Cattaneo equation.  Although
formally, the equation (\ref{eq8}) coincides with the known
expression for entropy production (see, for example, \cite{AK88}),
however, in the local nonequilibrium state entropy $S$ depends not
only on classical variables, but also on dissipative currents,
which, together with temperature and concentration, are considered
as independent variables, in our case $S = S(C, T, J_D^L) $. It
follows from the above that the chemical potentials $\mu^{LS} = -
T\partial S/\partial C_{LS}$ included in (\ref{eq8}) and (\ref{eq13})
are functions of the same variables.

Now let the interface move stationary at a velocity $V$ close to
$V_D$ ($V \lesssim V_D$) for which diffusionless solidification
takes place with $J_D^L = 0$ and the solute concentration in both
phases equal to the initial concentration in the melt $c_0$. Taking
into account the above,  one can write down for the chemical
potentials the following expansion
\begin{eqnarray}
  \mu_i^L(C_L, J_D^L) &=&  \mu_{leq,i}^L(C_l) - \alpha_i\frac{RT}{\rho V_{DI}M}J_D^L + ...\label{eq15} \\
\mu_i^S(C_S, J_D^S = 0)&=& \mu_{leq,i}^L(C_S).\label{eq16}
\end{eqnarray}
The terms in (\ref{eq15})-(\ref{eq16}) independent of $J_D^L$
represent the local equilibrium part of the chemical potential.  As
can be seen from (\ref{eq15}), in the first approximation $\mu$
depends linearly on $J_D^L$. In contrast to the bulk liquid, where
scalar functions can depend on the vector only through the scalar
product $J_D^2 = {\bm J}_D\cdot{\bm J}_D$, at the interface a
dependence on $J_D = {\bm n}\cdot{\bm J}_D$ is possible. The
coefficient at $J_D^L$ has been chosen such that $\alpha_i$ is a
dimensionless parameter of the order of unity, the sign of which
will be discussed later.

Using Eqs.(\ref{eq15})-(\ref{eq16}), let us find the explicit form
the Gibbs free energy change at the interface, $\Delta G_m$. It
should be noted that in the framework of  LNM the expression for it,
depending on $J_D^2$  only, has been earlier derived. (see, for
example, Ref.~\cite{GA18}). In fact, such expression can be obtained
from Eq.~(\ref{eq15}) if in this equation one retains second-order
terms proportional to $J_D^2$ and neglects the linear terms of the
current. A subsequent substitution of $\Delta\mu_i = \mu_i^S -
\mu_i^L$ in $\Delta G$ leads to Eq.~(2.14) from Ref.~\cite{GA18}).
However, taking into account that at $J_D^L \rightarrow 0$ the main
contribution to the chemical potential is just given  by the linear
terms in the current, further we will use Eq.~(\ref{eq15}) in the
linear approximation.

For the local equilibrium part of the chemical potential at $C$
close to $c_0$ in the linear approximation one can write
\begin{equation}\label{eq17}
    \mu_{leq}(C) =  \mu_{leq}(c_0) +  \frac{\partial \mu_{leq}(c_0)}{\partial
    c_0}(C - c_0) + ...
\end{equation}
Thus, in a state close to  diffusionless solidification the chemical
potentials at the interface can be represented as follows
\begin{eqnarray}
  \mu_i^L(C_L, J_D^L) &=&  \mu^{*L}_i + \frac{\partial \mu_{leq,i}^L(c_0)}{\partial
    c_0}(C_L - c_0) - \alpha_i\frac{RT}{\rho V_{DI}M}J_D^L + ...\label{eq18} \\
\mu_i^S(C_S)&=&  \mu^{*S}_i + \frac{\partial
\mu_{leq,i}^S(c_0)}{\partial
    c_0}(C_S - c_0) + ...,\label{eq19}
\end{eqnarray}
where
\begin{equation*}
    \mu^{*LS}_i = \mu_{leq,i}^{LS}(c_0) = \mu_i^{LS}(C_{LS} = c_0, J_D^{LS} = 0)
\end{equation*}
are the chemical potentials for the state of diffusionless
solidification.

In the case of dilute solution and using Henry's and
Raoult's laws (per unit mass) we have
\begin{eqnarray}
  \mu_{leq,A}^{LS}(c_0) &=&\mu_{0A}^{LS} + \frac{RT}{M}\ln(1 - c_0)\label{eq20}\\
 \mu_{leq,B}^{LS}(c_0)&=& \mu_{0B}^{LS} + \frac{RT}{M}\ln c_0,\label{eq21}
\end{eqnarray}
where $\mu_{0i}^{LS}$ is the standard chemical potential of species
$i$.  Calculating the derivatives of (\ref{eq20})-(\ref{eq21}) and
substituting them in (\ref{eq18})-(\ref{eq19}), one obtains for
thermodynamic forces
\begin{eqnarray}
 \Delta\mu_A &=& \mu_A^S - \mu_A^L = \Delta\mu_A^* + \frac{RT}{M}\frac{(C_L - C_S)}{(1 - c_0)} + \alpha_A\frac{RT}{\rho V_{DI}M}J_D^L  \label{eq22}\\
  \Delta\mu_B  &=& \mu_B^S - \mu_B^L = \Delta\mu_B^* - \frac{RT(C_L - C_S)}{c_0M} + \alpha_B\frac{RT}{\rho V_{DI}M}J_D^L, \label{eq23}
\end{eqnarray}
where
\begin{equation*}
  \Delta\mu_i^* = \mu^{*S}_i - \mu^{*L}_i
\end{equation*}
are the thermodynamic driving  forces for the state of diffusionless
solidification  satisfying, by virtue of (\ref{eq13}) at $C_{LS} =
c_0$, $V = V_D$ and $J_D^L = 0$, the equation
\begin{equation}\label{eq24}
 (1 - c_0)\Delta\mu_A^* - c_0\Delta\mu_B^* = - f(V_D) =
 \frac{RT}{M}\ln(1 - V_D/V_0).
\end{equation}
Now substituting Eqs.(\ref{eq22})-(\ref{eq23}) in Eq.(\ref{eq13})
and using the equality (\ref{eq24}), we obtain
\begin{eqnarray}
   (c_0 &-& C_S)(\Delta\mu_A^* - \Delta\mu_B^* ) + (1 -
   C_S)(RT/M)\left(\frac{C_L - C_S}{1 - c_0}  +  \frac{\alpha_AJ_D^L}{\rho V_{DI}} \right) \nonumber \\
 && -\, C_S(RT/M)\left(\frac{C_L - C_S}{ c_0} - \frac{\alpha_BJ_D^L}{\rho V_{DI}}\right)
  = (RT/M)\ln \frac{1 - V/V_0}{1 - V_D/V_0}\label{eq25}
\end{eqnarray}
Expression (\ref{eq25}) together with Eq.(\ref{eq14}) defines the required
boundary conditions in the high-speed region, $V \lesssim V_D$.

\section{The partition coefficient}
In the absence of diffusion in the solid phase, $C_S \approx c_0$,
and for a dilute solution, $1 - c_0 \approx 1$, Eq.(\ref{eq25})
reduces to
\begin{equation}\label{eq26}
    (\alpha_A + \alpha_Bc_0)J_D^L/\rho V_{DI} = \ln\frac{1 - V/V_0}{1 -
    V_D/V_0} > 0,
\end{equation}
if $V < V_D$.   
At $C_L > C_S$  from (\ref{eq14}) it follows that $J_D^L > 0$ and
$\alpha_A + \alpha_Bc_0 > 0$.  The last inequality is automatically
satisfied if $\alpha_A > 0$ and $\alpha_B > 0$.

For $C_S = c_0$, combination of Eq.(\ref{eq14}) with
Eq.~(\ref{eq26}) gives
\begin{equation}\label{eq27}
    C_L - c_0 = \frac{V_{DI}/V}{\alpha_A
+ \alpha_Bc_0 }\ln\frac{1 - V/V_0}{1 -
    V_D/V_0}.
\end{equation}
From Eq. (\ref{eq27}) we obtain the non-equilibrium solute partition
coefficient $k(V)$ ($V \lesssim V_D $) in the form
\begin{equation}\label{eq28}
    k(V) = \frac{c_0}{C_L} = \frac{c_0(\alpha_A
+ \alpha_Bc_0 )V/V_{DI}}{c_0(\alpha_A + \alpha_Bc_0 )V/V_{DI} +
\displaystyle{\ln\frac{1 - V/V_0}{1 - V_D/V_0}}}
\end{equation}

\begin{figure}[h]\centering
\includegraphics[width= 0.7\textwidth]{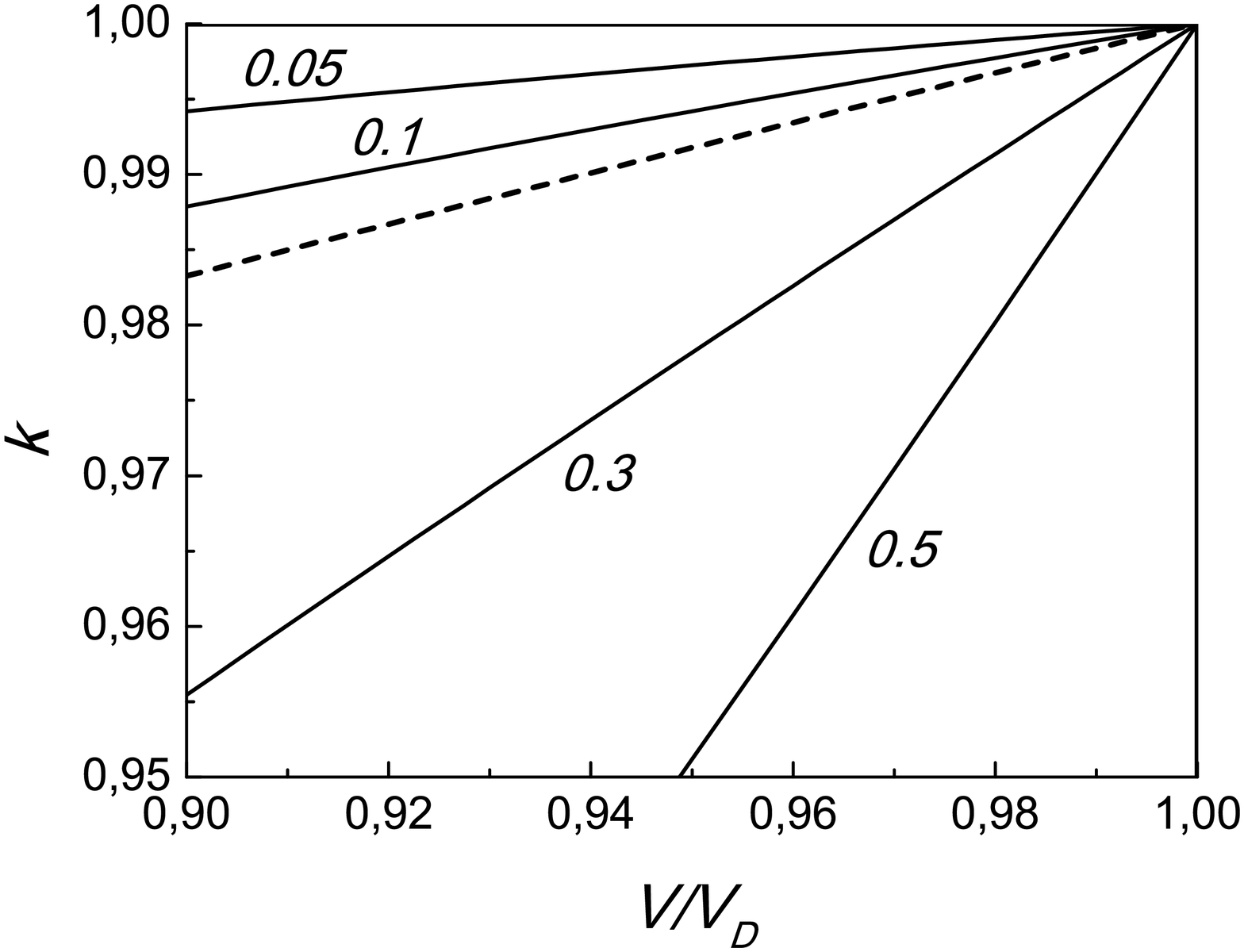}
      \parbox[t]{\textwidth}{\caption{ \label{fig1} The solute partition coefficients $k_1$ given by Eq.(\ref{eq28}) is shown by solid curves and the solute partition coefficient $k_2$ given by Eq.(\ref{eq1}) is shown by dotted curve both as functions of the ratio $V/V_D$. For the calculations we have chosen the following values $k_e = 0,1$;
       $c_0 = 0,1$; $V_D/V_{DI} = 1$; $\alpha_A = 1$,
       $\alpha_B = 0$, $V_D/V_0 = 0,05$; 0,1; 0,3; 0,5 (values are given at curves). }}
      \end{figure}

Figures 1 and 2 show the behavior of the
partition coefficients $k_1(V/V_D)$ and $k_2(V/V_D)$ calculated by (\ref{eq28}) and
(\ref{eq1}) at some values of the
parameters $V_D/V_{DI}$, $k_e$, $\alpha_A$, $\alpha_B$ and different
values of $V_D/V_0$ (indicated at curves). For a dilute solution,
the term $c_0^2$ in denominator of Eq. (\ref{eq28}) can be neglected
($\alpha_B = 0$). Parameter $\alpha_A$ (of the order of one) has
been taken to be equal to one, $\alpha_A =1$. As can be seen from
Figs. 1 and 2, the partition coefficients are quite close at relatively
small values of $V_D/V_0$, remaining, perhaps, within the
experimental errors. However as  $V_D/V_0$ increases, the
corresponding curves begin to differ markedly. At relatively large
$V_D/V_0$, the partition coefficient $k_1(V/V_D)$ changes rapidly enough
close to $V/V_D = 1$ compared to the coefficient $k_2(V/V_D)$.

\begin{figure}[h]\centering
    \includegraphics[width= 0.7\textwidth]{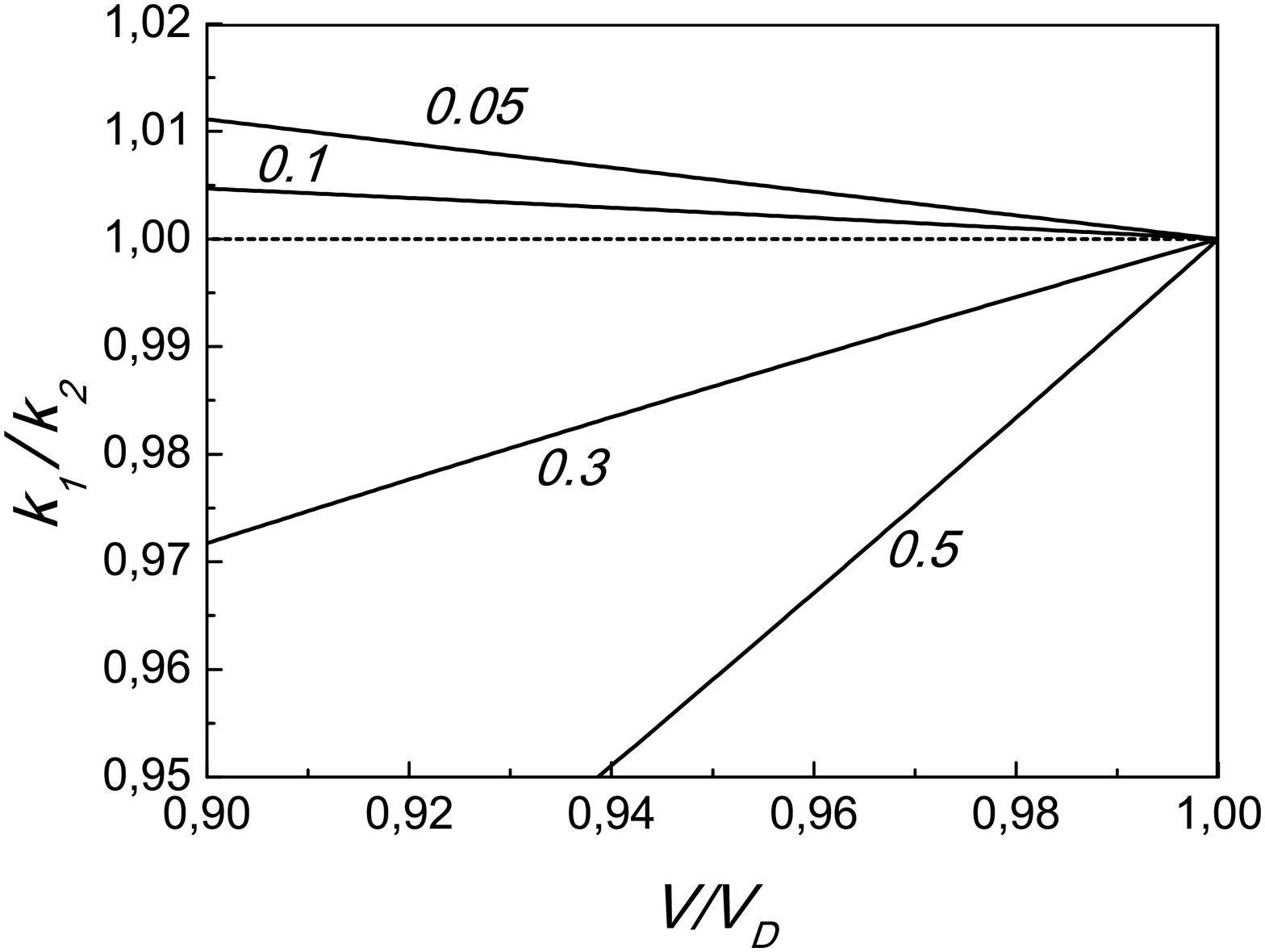}
      \parbox[t]{\textwidth}{\caption{ \label{fig2} The ratio of the solute partition coefficients $k_1/k_2$ {\it versus} $V/V_D$; $k_e = 0,1$;
       $c_0 = 0,1$; $V_D/V_{DI} = 1$; $\alpha_A = 1$,
       $\alpha_B = 0$, $V_D/V_0 = 0,05$; 0,1; 0,3; 0,5 (values are given at curves). }}
      \end{figure}

\section{Conclusion}
Unlike other models, LNM predicts the sharp transition to
diffusionless solidification and the complete solute trapping at a
finite interface velocity $V = V_D$. When the interface velocity
equal to or greater than the diffusion speed in the liquid $V_D$ the
solute atoms do not have time to diffuse into the bulk of the liquid
and are completely trapped by the interface with a concentration
equal to the initial melt concentration regardless
of the interfacial kinetics mechanism. This means that in the
high-speed region, $V\lesssim V_D$, the non-equilibrium partition
coefficient should mainly be dependent on the macroscopic boundary conditions at the interface.

Considering solidification of a binary mixture with a velocity $V =
V_D$ as a "reference state"  we have derived the boundary conditions
at the interface moving with a velocity $V \lesssim V_D$ and
determined the non-equilibrium partition coefficient (\ref{eq28}). A
comparison with the partition coefficient (\ref{eq1}) shows, that at relatively small values of $V_D/V_0$, where $V_0$ is
the upper boundary of the interface velocity, both coefficients show
similar behavior in the high-speed region. However with the increase
of $V_D/V_0$ their behavior sufficiently differs.


\begin{acknowledgement}
  The authors acknowledge the support by the European Space Agency (ESA) under
  research project MULTIPHAS (AO-2004), the German Aerospace Center (DLR) Space
  Management under contract No. 50WM1541 and also from the Russian Science
  Foundation under the project no. 16-11-10095.
\end{acknowledgement}

\section*{Author contribution statement}
All authors contributed equally to the present research article.



\end{document}